\documentclass[12pt]{article}
\usepackage{amssymb}
\usepackage{amsmath}
\usepackage{amsthm}

\newtheorem{theorem}{Theorem}[section]
\newtheorem{lemma}[theorem]{Lemma}
\newtheorem{corollary}[theorem]{Corollary}
\newtheorem{proposition}[theorem]{Proposition}
\numberwithin{equation}{section}

\begin{document}

\title{Unique solutions to boundary value problems in the cold plasma model}
\author{Thomas H. Otway\thanks{%
Department of Mathematics, Yeshiva University, New York, NY 10033
(otway@yu.edu)}
}
\date{}
\maketitle

\begin{abstract}
The unique existence of a weak solution to the homogeneous closed
Dirichlet problem on certain $D$-star-shaped domains is proven for a mixed
elliptic-hyperbolic equation. Equations of this kind arise in models
for electromagnetic wave propagation in cold plasma. A related class
of open boundary value problems is shown to possess strong
solutions.

\medskip

\noindent\textbf{Key words}. Equations of mixed elliptic-hyperbolic type, cold
plasma model, closed boundary value problem, symmetric positive
operator, Cinquini-Cibrario equation

\medskip

\noindent\textbf{AMS subject classifications}.  35M10, 35D30, 82D10

\end{abstract}

\section{Introduction}

Boundary value problems for mixed elliptic-hyperbolic equations
may be either \emph{open} or \emph{closed}. In the former case,
data are prescribed on a proper subset of the boundary whereas in
the latter case, data are prescribed on the entire boundary. It is
shown in Sec.\ 3 of \cite{MSW} that if $\kappa=1/2,$ the closed
Dirichlet problem is over-determined for the equation
\begin{equation}\label{coldpl1}
    \left(x-y^2\right)u_{xx}+u_{yy}+\kappa u_x=0
\end{equation}
on a typical domain, where $u\left(x,y\right)$ is required to be
twice-continuously differentiable on the domain. However, this
equation arises in a qualitative model for electromagnetic wave
propagation in an idealized cold plasma (\cite{W1}, eq.\ 81; see
also \cite{PF}, eq.\ (9)). Physical reasoning suggests that the
closed Dirichlet problem for (\ref{coldpl1}) should be well-posed in a suitable
function space, at least for some choice of lower order terms; so the result reported in \cite{MSW} would appear to represent a serious deficiency in the physical model. See
Sec.\ 1 of \cite{MSW} for a discussion, in which the problem of
formulating a closed Dirichlet-like problem that is well-posed in
an appropriate sense is characterized as an ``outstanding and
significant problem for the cold plasma model."

Except for the point at the origin, eq.\
(\ref{coldpl1}) can be transformed into an equation of Tricomi type;
see, \emph{e.g.}, Sec.\ 2.4.3 of \cite{O4}. Such equations have somewhat
more desirable analytic properties than eq.\ (\ref{coldpl1}).
However, both the physical and mathematical interest of eq.\
(\ref{coldpl1}) arise from the tangency of the resonance curve to a
flux line at the origin. This is the point at which plasma
heating might occur in the physical model, and a point which
appears to be singular in studies of the solutions to (\ref{coldpl1}); see
\cite{MSW}, \cite{PF}, and \cite{W1}. Thus it is not generally useful to transform
eq.\ (\ref{coldpl1}) into an equation of Tricomi type.

Using methods introduced by Lupo, Morawetz, and Payne \cite{LMP},
\cite{LMPE} for equations of Tricomi type, we show in Sec.\ 2 the
weak existence of a unique solution to a homogeneous closed
Dirichlet problem for the formally self-adjoint ($\kappa=1$) case of eq.\
(\ref{coldpl1}). This extends a recent result \cite{O2} in which
the existence of solutions having various degrees of smoothness
was shown in certain cases to which uniqueness proofs did not seem
to apply. At the same time, it extends the unique-existence
arguments in \cite{LMP} to an equation which is not of Tricomi
type.

Another well known problem in elliptic-hyperbolic theory is the
determination of natural conditions for boundary geometry; see the
discussions in \cite{BG}, \cite{Mo1}, \cite{Mo2}, \cite{P1},
\cite{P2}, and \cite{Pi}. Heuristic approaches to determining
boundary geometry tend to focus on physical \cite{Ma} or geometric
\cite{O3} analogies for the specific equation under study. In his
theory of symmetric positive systems \cite{F}, Friedrichs proposed
intrinsic mathematical criteria for the well-posedness, or
\emph{admissibility}, of boundary conditions. But Friedrichs'
conditions are also tied to the specifics of the particular
symmetric positive equation under study and are algebraic rather
than explicitly geometric. We will require boundary arcs to be
starlike with respect to an appropriate vector field. This
approach to boundary geometry was introduced by Lupo and Payne
\cite{LP2}. We note that algebraic conditions in certain very old
results can be reinterpreted as the requirement of a starlike
boundary; see, \emph{e.g.}, \cite{K}. Our results provide further
evidence that domain boundaries which are starlike in this
generalized sense are natural for elliptic-hyperbolic boundary
value problems.

In Sec.\ 3 we investigate the solvability of open boundary value
problems for a class of symmetric positive systems on domains having
starlike boundaries. This class includes an equation originally
studied by Cinquini-Cibrario in the early 1930s. The equation
arises in the so-cold ``slab" model of zero-temperature plasma and
in models of low-temperature atmospheric and space plasmas (Sec.\ 3.1).

The boundary conditions in Sec.\ 3 are
\emph{mixed} in the sense that a Dirichlet condition is placed on
part of the boundary and a Neumann condition is placed on another
part. However, our methods also apply to the case in which either
a Dirichlet or a Neumann condition is imposed over the entire
elliptic boundary. Because the boundary value problems considered
in Sec.\ 3 are open, the results of that section may be less
interesting physically than those of Sec.\ 2. But open boundary
conditions can be expected to imply more smoothness on the part of
solutions than is obtained from closed boundary conditions, and we provide
conditions for the existence of solutions which are strong in the sense of
Friedrichs. Conditions for the existence of strong solutions to
elliptic-hyperbolic boundary value problems were also discussed in
Sec.\ 3 of \cite{O2}, but briefly and inadequately. Section 3 of
this report revises and extends (to the open case of mixed and
Neumann problems) the treatment of strong solutions in \cite{O2}.
The existence question for weak solutions to open boundary value
problems for equations of the form (\ref{coldpl1}) was considered
in \cite{O1} and \cite{Y}.

\subsection{Remarks on the physical model}

In the cold plasma model, the plasma temperature is assumed to be
zero in order to neglect the fluid properties of the medium, which
is treated as a linear dielectric. In the case of wave propagation through an underlying static
medium having axisymmetric geometry, equations of the form
(\ref{coldpl1}) model the tangency of a flux surface to a
resonance surface. At the  point of tangency, plasma heating might
occur even in the cold plasma model \cite{W1}. In two dimensions, flux
surfaces (level sets of the magnetic flux function) can be
represented by the lines $x=\mbox{const.},$ and a resonance
surface (frequencies at which the field equations change from
elliptic to hyperbolic type) by the curve $x=y^2.$ In such cases a plasma heating zone could
lie at the origin of coordinates. This conjecture is supported by
numerical \cite{MSW} and classical \cite{PF} analysis which
suggests that the origin is a singular point of eq.\
(\ref{coldpl1}).

The main physical implication of our result is that, although the Dirichlet problem for the cold plasma model is manifestly ill-posed in the classical sense, there is a weak sense in which the closed Dirichlet problem is no longer over-determined:  a unique weak solution to the closed Dirichlet problem exists in an appropriately weighted function space. The weight function vanishes on the resonance curve. In Corollary 6 of \cite{O2} an existence theorem, without uniqueness, was demonstrated in a weighted function space in which the weight function vanished on the line $y=0.$ Thus in each case the weight function vanishes at the origin of coordinates. These results support the physical conjecture of a heating zone at that point.

For discussions of the physical context of eq.\
(\ref{coldpl1}), see Sec.\ V of \cite{W1} and Sec.\ 4 of
\cite{W2}, in which eq.\ (\ref{coldpl1}) with $\kappa=0$ is proposed
as a qualitative model for erratic heating effects by lower hybrid
waves in the plasma. See also \cite{PF}, in which a model for
electrostatic waves in a cold anisotropic plasma with a
two-dimensional inhomogeneity yields, by a formal derivation, an
equation for the field potential which is similar to
(\ref{coldpl1}). Briefly, the derivation proceeds as follows: at zero temperature, the field equations reduce to Maxwell's equations, which are written for the electric displacement vector $\mathbf{D}.$ The components of $\mathbf{D}$ are written in terms of the dielectric tensor for the medium. The coordinate system is chosen so that, in two dimensions, the $y$-axis is collinear to a longitudinally applied magnetic field. In that case the dielectric tensor assumes a particular form which leads, when the resonance curve is tangent to the flux line at the origin, to an equation having the form considered here. Precisely, the equation derived in \cite{PF} is
eq.\ (\ref{PFeq}) of Remark \emph{iii)}, Sec.\ 3, below, with
particular choices of $u_1,$ $u_2,$ $\sigma(y),$ $\kappa_1,$ and
$\kappa_2.$

The derivation of the governing equation for the case of fully electromagnetic waves is much more lengthy, and occupies much of the memoir \cite{W1}. In that case, equations having the form (\ref{coldpl1}) provide only a qualitative description of the physical model. Section 2.5 of \cite{O4} is devoted to a brief derivation of eq.\ (\ref{coldpl1}) for electromagnetic waves.

The classic text for wave propagation in plasma is \cite{St}; see
also \cite{G}. See Ch.\ 2 of \cite{Sw} for a concise introduction to
the physics of electromagnetic waves in cold plasma, and \cite{O4}
for a recent review of mathematical aspects of the cold plasma
model.

In the sequel we assume that $\Omega$ is an open, bounded, connected
domain of $\mathbb{R}^2$ having at least piecewise continuously differentiable
boundary with counterclockwise orientation. For nontriviality we
require that $\Omega$ contain an arc of the resonance curve
$x=y^2$ ($x = \sigma(y)$ in Sec.\ 3). Additional conditions will be placed on the
domain where required.

\section{Weak solutions to closed boundary value problems}

Following Sec.\ 3 of \cite{LMP} we define, for a given $C^1$
function $K\left(x,y\right),$ the space
$L^2\left(\Omega;|K|\right)$ and its dual. These spaces consist,
respectively, of functions $u$ for which the norm
\[
||u||_{L^2\left(\Omega;|K|\right)}=\left(\int\int_\Omega|K|u^2dxdy\right)^{1/2}
\]
is finite, and functions $u\in L^2\left(\Omega\right)$ for which
the norm
\[
||u||_{L^2\left(\Omega;|K|^{-1}\right)}=\left(\int\int_\Omega|K|^{-1}u^2dxdy\right)^{1/2}
\]
is finite. Analogously, we define the space $H^1(\Omega; K)$ to
be the completion of $C_0^\infty(\Omega)$ with respect to the norm
\begin{equation}\label{H101}
    ||u||_{H^1(\Omega; K)}=\left[\int\int_{\Omega}
\left(|K|u_x^2+u_y^2+u^2\right)\,dxdy\right]^{1/2},
\end{equation}
and introduce the space $H^1_0(\Omega; K)$ as a closure in this space. Using a weighted Poincar\'e inequality to absorb the zeroth-order
term, we write the $H^1_0(\Omega; K)$-norm in the form
\begin{equation}\label{H102}
    ||u||_{H^1_0(\Omega; K)}=\left[\int\int_\Omega
\left(|K|u_x^2+u_y^2\right)\,dxdy\right]^{1/2}.
\end{equation}
The dual space $H^{-1}\left(\Omega; K\right)$ is defined via the negative norm
\[
\vert\vert w\vert\vert_{H^{-1}\left(\Omega;K\right)}=\sup_{0\ne\varphi\in C_0^\infty\left(\Omega\right)}\frac{\left\vert\langle w,\varphi\rangle\right\vert}{\vert\vert\varphi\vert\vert_{H_0^1\left(\Omega,K\right)}},
\]
where $\langle\,\,,\,\,\rangle$ is the Lax duality bracket.

Various lower-order terms have been associated to eq.\
(\ref{coldpl1}) in the literature on the cold plasma model;
only the higher-order terms contribute to resonance. These lower-order terms were considered explicitly in Sec.\ 2 of \cite{O2}, in which we showed the existence of $L^2$ solutions to a closed Dirichlet problem for eq.\ (\ref{coldpl1}). Here we consider only the formally self-adjoint case
\begin{equation}\label{coldpl2}
    Lu \equiv \left[K\left(x,y\right)u_x\right]_x+u_{yy}=f\left(x,y\right),
\end{equation}
for the type-change function $K\left(x,y\right)=x-y^2.$ As in \cite{O2}, the
inhomogeneous term $f\left(x,y\right)$ is assumed known. For this special case of lower-order terms, we are able to show the existence of solutions to the closed Dirichlet problem having much higher regularity than was the case in \cite{O2}.

The forcing function $f$ in eq.\ (\ref{coldpl2}) (and also in eq.\ (\ref{sys1}), below) could arise physically under appropriate conditions on the charge density. In addition, the presence of $f$ provides mathematical generality, avoids trivial solutions, and allows the conversion of solutions to the Dirichlet problem with homogeneous boundary conditions into solutions to a problem for the homogeneous equation with inhomogeneous boundary conditions. Finally, analysis of the inhomogeneous equation is useful when, as in our case, one expects singularities in the homogeneous equation.

In accordance with standard terminology, we will often refer to
the curve $K=0$ on which eq.\ (\ref{coldpl2}) changes type as the
\emph{sonic curve.} This terminology is borrowed from fluid
dynamics; in the context of the cold plasma model, the sonic
transition occurs at a resonance frequency.

Following Lupo, Morawetz, and Payne \cite{LMP}, we define a
\emph{weak solution} of eq.\ (\ref{coldpl2}) on $\Omega,$ with
boundary condition
\begin{equation}\label{boundary}
    u(x,y)=0\,\forall\, (x,y)\in\partial\Omega,
\end{equation}
to be a function $u\in H^1_0(\Omega;K)$ such that $\forall \xi \in
H^1_0(\Omega;K)$ we have
\[
\langle Lu, \xi\rangle\equiv
-\int\int_\Omega\left(Ku_x\xi_x+u_y\xi_y\right)dxdy =
\langle f,\xi\rangle,
\]
where $\langle\,\,,\,\,\rangle$ is duality pairing between $H^1_0\left(\Omega,K\right)$ and $H^{-1}\left(\Omega,K\right).$ In
this case the existence of a weak solution is equivalent to the
existence of a sequence $u_n\in C_0^\infty(\Omega)$ such that
\[
||u_n-u||_{H_0^1\left(\Omega;K\right)}\rightarrow 0 \mbox{ and }
||Lu_n-f||_{H^{-1}\left(\Omega;K\right)}\rightarrow 0
\]
as $n$ tends to infinity.

Following Lupo and Payne (Sec.\ 2 of \cite{LP2}), we consider a
one-parameter family $\psi_\lambda\left(x,y\right)$ of
inhomogeneous dilations given by
\[
\psi_\lambda\left(x,y\right)=\left(\lambda^{-\alpha}x,\lambda^{-\beta}y\right),
\]
where $\alpha, \beta,\lambda \in \mathbb{R}^+,$ and the associated
family of operators
\[
\Psi_\lambda u = u\circ \psi_\lambda \equiv u_\lambda.
\]
Denote by $D$ the vector field

\begin{equation}\label{vector}
    Du = \left[\frac{d}{d\lambda}u_\lambda\right]_{|\lambda=1}=-\alpha
x
\partial_x-\beta y\partial_y.
\end{equation}
An open set $\Omega \subseteq \mathbb{R}^2$ is said to be
\emph{star-shaped} with respect to the flow of $D$ if
$\forall\left(x_0,y_0\right)\in \overline\Omega$ and each
$t\in\left[0,\infty\right]$ we have
$F_t\left(x_0,y_0\right)\subset \overline\Omega,$ where
\[
F_t\left(x_0,y_0\right)=\left(x(t),y(t)\right)=\left(x_0e^{-\alpha
t},y_0e^{-\beta t}\right).
\]
If a domain is star-shaped with respect to a vector field $D,$
then it is possible to ``float" from any point of the domain to
the origin along the flow lines of the vector field. If these flow
lines are straight lines through the origin
$\left(\alpha=\beta\right),$ then we recover the conventional
notion of a star-shaped domain. By an appropriate translation, the
origin can be replaced by any point $\left(x_s,y_s\right)$ in the
plane as a source of the flow. In that case we obtain a translated
function $\tilde F_t$ for which
\[
\lim_{t\rightarrow\infty}\tilde
F_t\left(x_0,y_0\right)=\left(x_s,y_s\right)\,\,\forall\left(x_0,y_0\right)\in\overline\Omega.
\]
Moreover, whenever a domain is star-shaped with respect to the
flow of a vector field satisfying (\ref{vector}), the domain
boundary will be \emph{starlike} in the sense that
\[
\left(\alpha x,\beta
y\right)\cdot\hat{\mathbf{n}}\left(x,y\right)\geq 0,
\]
where $\hat{\mathbf{n}}$ is the outward-pointing normal vector on
the boundary $\partial\Omega.$ See Lemma 2.2 of \cite{LP2}. In
equivalent notation, given a vector field $V=-\left(b,c\right)$
and a boundary arc $\Gamma$ which is starlike with respect to $V,$
the inequality
\begin{equation}\label{starlike1}
    bn_1+cn_2 \geq 0
\end{equation}
is satisfied on $\Gamma.$

\subsection{An auxiliary problem}

We employ the \emph{dual variational method}, an integral variant
of the $abc$ method, introduced by Didenko \cite{D} and developed
by Lupo and Payne \cite{LP1}. (Note that the term ``dual variational" also appears in elliptic variational theory, in which it means something entirely different.) Denote by $v$ a solution to
the boundary value problem
\begin{equation}\label{cauchy}
    Hv=u \mbox{ in } \Omega
\end{equation}
for $u\in C_0^\infty(\Omega),$ with $v$ vanishing on
$\partial\Omega\backslash \lbrace\left(0,0\right)\rbrace,$
\begin{equation}\label{limit}
    \lim_{\left(x,y\right)\rightarrow\left(0,0\right)}v\left(x,y\right)=0,
\end{equation}
and
\begin{equation}\label{oper}
    Hv = av + bv_x+cv_y.
\end{equation}
Assume that $\Omega$ is star-shaped with respect to the flow of
the vector field $V=-\left(b,c\right);$ $b=mx$ and $c=\mu y;$
$\mu$ and $m$ are positive constants and $a$ is a negative
constant; the point $\left(x,y\right)=\left(0,0\right)$ lies on
$\partial\Omega.$ Step 1 in the proof of Lemma 3.3
in \cite{LMP}, which treats the harder case of a non-differentiable
coefficient, demonstrates that $v$ satisfying (\ref{cauchy})-(\ref{oper}) exists and lies in the space $C^0\left(\overline{\Omega}\right)\cap H_0^1\left(\Omega;K\right).$ The proof is straightforward in our case, except perhaps for the justification of (\ref{limit}), which we outline:

Because $\Omega$ is star-shaped with respect to
$-\left(b,c\right),$ any flow line entering the interior of
$\Omega$ from $\partial\Omega$ will remain in $\overline\Omega.$
So the method of characteristics yields a smooth solution to
(\ref{cauchy}), (\ref{oper}), with a possible singularity at the
origin. It remains only to analyze the limiting behavior of the
solution near the origin. Because the origin lies on the boundary
and $u$ has compact support, we can restrict our attention to an
$\varepsilon$-neighborhood $N_\varepsilon$ of the boundary,
\[
N_\varepsilon\left(\partial\Omega\right) =
\left\{\left(x,y\right)\in\overline\Omega |
\mbox{dist}\left(\left(x,y\right),\partial\Omega\right)
\leq\varepsilon\right\},
\]
where $\varepsilon$ is so small that $N_\varepsilon$ lies in
$\overline\Omega$ but outside the support of $u$ in $\Omega.$ In
this subdomain we solve the Cauchy problem for the homogeneous
equation
\begin{equation}\label{homog}
    mxv_x+\mu yv_y=|a| v.
\end{equation}
The characteristic equation for (\ref{homog}), given by  $mxdy =
\mu y dx,$ can be integrated to yield
\begin{equation}\label{charac}
    y = cx^{\mu/m},
\end{equation}
where $c$ is a constant. The method of characteristics thus yields
a general solution in $N_\varepsilon$ having the form
\begin{equation}\label{gen}
    v\left(x,y\right) = \varphi\left(\frac{x^\mu}{y^m}\right)y^{|a|/\mu},
\end{equation}
where $\varphi$ is an arbitrary $C^1$ function that may be
prescribed along a non-characteristic curve in $N_\varepsilon.$
Following the evolution of the solution (\ref{gen}) along the
curves (\ref{charac}), we obtain (\ref{limit}) as in eq.\ (3.13)
of \cite{LMP}.

Following Sec.\ 2 of \cite{D} (see also the Appendix to
\cite{LP1}), the dependence of the solution $v$ to the problem
(\ref{cauchy})-(\ref{oper}) on the forcing function $u$ is
represented by an operator $\mathcal{I},$ writing
$v=\mathcal{I}u.$ We have the integral identities
\begin{equation}\label{intiden}
    \left(\mathcal{I}u,Lu\right)=\left(v,Lu\right)=\left(v,LHv\right).
\end{equation}
A good choice of the coefficients $a,$ $m,$ and $\mu$ in the operator $H$
on the right-hand side of this identity will allow us to derive an energy
inequality, which will be used to prove weak existence via the
Riesz Representation Theorem; see Ch.\ 2 of \cite{B} for a general
treatment of such ``projection" arguments.

\subsection{Main result}

The following is a small but crucial extension of \cite{O2}, Theorem
5.

\begin{lemma}\label{Lemma1} Suppose that $x$ is non-negative on
$\overline{\Omega}$ and that the origin of coordinates lies on
$\partial\Omega.$ Let $\Omega$ be star-shaped with respect to the
flow of the vector field $V=-\left(b,c\right)$ for $b=mx$ and $c=\mu
y,$ where $m$ and $\mu$ are positive constants and $m$ exceeds
$3\mu.$ Then there exists a positive constant $C$ for which the
inequality
\[
    ||u||_{L^2\left(\Omega;|K|\right)}\leq
C||Lu||_{H^{-1}\left(\Omega;K\right)}
\]
holds for every $u \in C^\infty_0\left(\Omega\right),$ where
$K(x,y)=x-y^2$ and $L$ is defined by (\ref{coldpl2}).
\end{lemma}

\emph{Proof.} Let $v$ satisfy eqs.\ (\ref{cauchy})-(\ref{oper}) on
$\Omega$ for $a=-M,$ where $M$ is a positive number satisfying
\[
M=\frac{m-3\mu}{2}-\delta
\]
for some sufficiently small positive number $\delta.$ Integrate
the identities (\ref{intiden}) by parts, using Prop.\ 12
of \cite{O2} and the compact support of $u.$ We have
\begin{eqnarray}
\int\int_\Omega v\cdot LHv\,dxdy =
\frac{1}{2}\oint_{\partial\Omega}
\left(Kv_x^2+v_y^2\right)\left(cdx-bdy\right) \nonumber\\
+ \int\int_\Omega\alpha v_x^2 + \gamma v_y^2
     dxdy,\label{prop12}
     \end{eqnarray}
where
\[
\alpha =
K\left(\frac{c_y-b_x}{2}-a\right)+\frac{1}{2}b+\frac{1}{2}K_yc
\]
\[
=\left(\frac{m}{2}-\mu-\delta\right)x+\delta y^2
\]
and
\[
\gamma = -a - \frac{c_y}{2}+\frac{b_x}{2}
=M-\frac{\mu-m}{2}=m-2\mu-\delta>\mu - \delta.
\]
On the \emph{elliptic} region $\Omega^+,$ $K>0$ and
\[
\left(\frac{m}{2}-\mu-\delta\right)x>\left(\frac{\mu}{2}-\delta\right)x\geq
\delta x
\]
provided we choose $\delta$ so small that $\mu/4\geq\delta.$ Then
on $\Omega^+,$
\[
\alpha \geq \delta\left(x+y^2\right)\geq
\delta\left(x-y^2\right)=\delta K=\delta|K|.
\]
On the \emph{hyperbolic} region $\Omega^-,$ $K<0$ and
\[
\alpha =
\left(\frac{m}{2}-\mu\right)x+\delta\left(y^2-x\right)\geq
\frac{\mu}{2}x+\delta\left(-K\right)\geq \delta|K|.
\]

We show that the integrand of the boundary integral in (\ref{prop12}) is non-negative: Because $v\in C^0\left(\overline{\Omega}\right)\cap H^1_0\left(\Omega;K\right),$
\[
\vert K\vert v_x^2+v_y^2=0
\]
in a sufficiently small neighborhood of $\partial\Omega.$ This implies that $v_x^2=v_y^2=0$ in a sufficiently small neighborhood of $\partial\Omega\backslash\left\{K=0\right\}.$ That in turn implies that
\[
Kv_x^2+v_y^2=0
\]
in a sufficiently small neighborhood of $\partial\Omega\backslash\left\{ K=0\right\}.$ On the curve $K=0,$ we have $x=y^2$ and $dx=2ydy,$ implying that
\[
cdx-bdy=\left(2\mu y^2-mx\right)dy=\left(2\mu-m\right)xdy.
\]
But $m>3\mu>2\mu,$ $x\geq 0$ on $\overline{\Omega},$ and $dy\leq 0$ on $K;$ so on the resonance curve,
\[
v_y^2\left(cdx-bdy\right)\geq 0.
\]
Thus the integrand of the boundary integral in (\ref{prop12}) is bounded below by zero.

We find that if $\delta$ is sufficiently small relative to $\mu,$
then
\begin{equation}\label{fin1}
    \left(v,LHv\right) \geq \delta\int\int_\Omega\left( |K| v_x^2 +
    v_y^2\right)
     dxdy.
\end{equation}
The upper estimate is immediate, as
\begin{equation}\label{fin2}
    \left(v,LHv\right) =
    \left(v,Lu\right)\leq\left\|v\right\|_{H_0^1\left(\Omega;K\right)}\left\|Lu\right\|_{H^{-1}\left(\Omega;K\right)}.
\end{equation}
Combining (\ref{fin1}) and (\ref{fin2}), we obtain
\begin{equation}\label{pen}
    \left\|v\right\|_{H_0^1\left(\Omega;K\right)}\leq
C\left\|Lu\right\|_{H^{-1}\left(\Omega;K\right)}.
\end{equation}
The assertion of Lemma \ref{Lemma1} now follows from (\ref{cauchy})
by the continuity of $H$ as a map from $H_0^1\left(\Omega;K\right)$
into $L^2\left(\Omega;|K|\right).$ This completes the proof of Lemma
\ref{Lemma1}.

\begin{theorem}\label{Theorem2} Let $\Omega$ be star-shaped with
respect to the flow of the vector field $-V=\left(mx, \mu y\right),$
where $m$ and $\mu$ are defined as in Lemma \ref{Lemma1}. Suppose
that $x$ is nonnegative on $\overline{\Omega}$ and that the origin of
coordinates lies on $\partial\Omega.$ Then for every $f \in
L^2\left(\Omega;|K|^{-1}\right)$ there is a unique weak solution
$u\in H_0^1\left(\Omega; K\right)$ to the Dirichlet problem
(\ref{coldpl2}), (\ref{boundary}) where $K=x-y^2.$
\end{theorem}

\emph{Proof}. The proof follows the outline of the arguments in
\cite{LMP}, Sec.\ 3. Defining a linear functional $J_f$ by the
formula
\[
J_f\left(L\xi\right)=\left(f,\xi\right),\,\xi\in
C_0^\infty\left(\Omega\right),
\]
we estimate
\[
|J_f\left(L\xi\right)|\leq||f||_{L^2\left(\Omega;|K|^{-1}\right)}||\xi||_{L^2\left(\Omega;|K|\right)}
\leq
C||f||_{L^2\left(\Omega;|K|^{-1}\right)}||L\xi||_{H^{-1}\left(\Omega;K\right)},
\]
using Lemma \ref{Lemma1}. Thus $J_f$ is a bounded linear functional
on the subspace of $H^{-1}\left(\Omega;K\right)$ consisting of
elements having the form $L\xi$ with $\xi\in
C_0^\infty\left(\Omega\right).$ Extending $J_f$ to the closure of
this subspace by Hahn-Banach arguments, the Riesz Representation
Theorem guarantees the existence of an element $u\in
H^1_0\left(\Omega;K\right)$ for which
\[
\left\langle u,L\xi\right\rangle=\left(f,\xi\right),
\]
where $\xi\in H^1_0\left(\Omega; K\right).$ There exists a unique,
continuous, self-adjoint extension
$L:H^1_0\left(\Omega;K\right)\rightarrow
H^{-1}\left(\Omega;K\right).$ As a result, if a sequence $u_n$ of
smooth, compactly supported functions approximates $u$ in the norm
$H^1_0\left(\Omega;K\right),$ then $Lu_n$ converges in norm to an
element $\tilde f$ of $H^{-1}\left(\Omega;K\right).$ Taking the
limit
\[
\lim_{n\rightarrow\infty}\left\langle
u-u_n,L\xi\right\rangle=\left(f-\tilde f,\xi\right),
\]
we conclude that, because the left-hand side vanishes for all
$\xi\in H^1_0\left(\Omega;K\right),$ the right-hand side must vanish
as well. This proves the existence of a weak solution. Taking the
difference of two weak solutions, we find that this difference is
zero in $H^1_0\left(\Omega;K\right)$ by Lemma \ref{Lemma1}, the
linearity of $L,$ and the weighted Poincar\'e inequality \cite{LMP}.
This completes the proof of Theorem \ref{Theorem2}.

\bigskip

The unique-existence proofs of this section use estimates similar
to those used in the proofs in Sec.\ 4 of \cite{O2} for existence
alone. But the likelihood that the estimates would turn out to be
similar appeared to be small on the basis of previous literature,
and is rather surprising. In Sec.\ 5.1 of \cite{O2} it is shown
that the estimates used to prove weak existence do not extend in
an obvious way to proofs of uniqueness for the case $K=x,$ in
which the resonance curve is collinear with the flux line. Based
on the physical discussion on p.\ 42 of \cite{W1}, the collinear
case would appear to be simpler than the case treated here, in
which the two curves are tangent at an isolated point. But the
simplicity of the case $K=x$ arises largely from the fact that in
this case the plasma behaves like a perpendicularly stratified
medium, in which wave motion satisfies an ordinary differential
equation; \emph{c.f.} \cite{LM}. The cautionary example of
\cite{O2}, Sec.\ 5.1, which suggests the difficulty of modifying
the weak-existence methods to prove uniqueness in the case $K=x,$
happens to fail in the case $K= x-y^2.$ As we have shown in this
section, a modification of the weak existence estimates will in
fact lead to a uniqueness proof for weak solutions to
(\ref{coldpl2}), (\ref{boundary}) for our choice of $K.$ The
case $K=x$ remains interesting from several points of view, and we
will return to its study in Sec.\ 3.1.

The restriction that the points of $\Omega$ may not lie in the negative half-plane corresponds a requirement that boundary conditions be placed on one side of the flux line. As the resonance frequency is naturally restricted to the same side of that line, the requirement seems to be compatible with the physical model.

\section{Strong solutions to open boundary value problems}

In this section we seek conditions sufficient for the existence of
strong solutions in the cold plasma model. The conditions that we
find will turn out to be extremely restrictive, but they are satisfied in a well known special case.

Consider a system of the form
\begin{equation}\label{sys1}
L\mathbf{u}=\mathbf{f}
\end{equation}
for an unknown vector
\[
\mathbf{u}=\left( u_{1}\left( x,y\right) ,u_{2}\left( x,y\right)
\right),
\]
and a known vector
\[
\mathbf{f}=\left( f_{1}\left( x,y\right) ,f_{2}\left( x,y\right)
\right),
\]
where $\left( x,y\right) \in \Omega \subset \mathbb{R}^2.$ The
operator $L$ satisfies
\begin{equation}\label{sys2}
    \left( L\mathbf{u}\right) _{1}=K\left(x,y\right)u_{1x} + u_{2y}+\mbox{ zeroth-order terms},
\end{equation}

\begin{equation}\label{sys3}
    \left( L\mathbf{u}\right) _{2}=u_{1y}-u_{2x}.
\end{equation}
As in the preceding section, $K\left(x,y\right)$ is continuously
differentiable, negative on $\Omega^-,$ positive on $\Omega^+,$
and zero on a parabolic region (the resonance curve) separating the
elliptic and hyperbolic regions. If
$\left(f_1,f_2\right)=\left(f,0\right),$ the components of the
vector $\mathbf{u}$ are continuously differentiable, and
$u_1=u_x,$ $u_2=u_y$ for some twice-differentiable function
$u(x,y),$ then the first-order system (\ref{sys1})-(\ref{sys3})
reduces to a second-order scalar equation such as (\ref{coldpl2}).
Because the emphasis in this section is on the form of the
boundary conditions, the presence or absence of zeroth-order terms
will not affect the arguments provided the resulting system is
symmetric positive.

We say that a vector $\mathbf{u}=(u_1,u_2)$ is in $L^2$ if each of
its components is square-integrable. Such an object is a
\textit{strong solution} of an operator equation of the form
(\ref{sys1}), with given boundary conditions, if there exists a
sequence $\mathbf{u}^{\nu }$ of continuously differentiable
vectors, satisfying the boundary conditions, for which
$\mathbf{u}^{\nu }$ converges to $\mathbf{u}$ in $L^2$ and
$L\mathbf{u}^{\nu }$ converges to $\mathbf{f}$ in $L^2.$

Sufficient conditions for a vector to be a strong solution were
formulated by Friedrichs \cite{F}. An operator $L$ associated to
an equation of the form
\begin{equation}\label{matrixeq}
L\mathbf{u}=A^1\mathbf{u}_x+A^2\mathbf{u}_y+B\mathbf{u},
\end{equation}
where $A^1$, $A^2$, and $B$ are matrices, is said to be
\emph{symmetric positive} if the matrices $A^1$ and $A^2$ are
symmetric and the matrix
\[
    Q \equiv B^* - \frac{1}{2}\left( A_x^1 + A_y^2\right)
\]
is positive-definite, where $B^*$ is the symmetrization of the
matrix $B:$
\[
B^* = \frac{1}{2} \left(B+B^T\right).
\]
The differential equation associated to a symmetric positive
operator is also said to be symmetric positive.

Boundary conditions for a symmetric positive equation can be given
in terms of a matrix
\begin{equation}\label{bta}
    \beta = n_1A_{|\partial \Omega}^1 + n_2A_{|\partial \Omega}^2,
\end{equation}
where $\left(n_1,n_2\right)$ are the components of the
outward-pointing normal vector on $\partial \Omega.$ The boundary
is assumed to be twice-continuously differentiable. Denote by
$\mathcal{V}$ the vector space identified with the range of
$\mathbf{u}$ in the sense that, considered as a mapping, we have
$\mathbf{u}:\Omega \cup\partial \Omega \rightarrow \mathcal{V}.$
Let $\mathcal{N}(\tilde x, \tilde y),$ $(\tilde x,\tilde
y)\in\partial\Omega,$ be a linear subspace of $\mathcal{V}$ and
let $\mathcal{N}(\tilde x, \tilde y)$ depend smoothly on $\tilde
x$ and $\tilde y.$ A boundary condition $u \in \mathcal{N}$ is
\emph{admissible} if $\mathcal{N}$ is a maximal subspace of
$\mathcal{V}$ with respect to non-negativity of the quadratic form $(\mathbf{u},\beta
\mathbf{u})$ on the boundary.

A set of sufficient conditions for admissibility is the existence
of a decomposition (\cite{F}, Sec.\ 5)
\begin{equation}\label{sum}
    \beta = \beta_++\beta_-,
\end{equation}
for which: the direct sum of the null spaces for $\beta_+$ and
$\beta_-$ spans the restriction of $\mathcal{V}$ to the boundary;
the ranges $\mathfrak{R}_\pm$ of $\beta_\pm$ have only the vector
$\mathbf{u}=0$ in common; and the matrix $\mu=\beta_+-\beta_-$
satisfies
\begin{equation}\label{difference}
    \mu^\ast = \frac{\mu + \mu^T}{2} \geq 0.
\end{equation}
These conditions imply that the boundary condition
\begin{equation}\label{bndrycon}
    \beta_-\mathbf{u}=0 \,\, \mbox{on $\partial \Omega$}
\end{equation}
is admissible for eq.\ (\ref{sys1}) and the boundary condition
\begin{equation}\label{adj}
    \mathbf{w}^T \beta_+^T =0 \,\, \mbox{on $\partial \Omega$}
\end{equation}
is admissible for the adjoint problem
\[
L^\ast \mathbf{w}=\mathbf{g} \,\, \mbox{in $\Omega.$}
\]
The linearity of the operator $L$ and the admissibility conditions
on the matrices $\beta_\pm$ imply that both problems possess
unique, strong solutions.

Boundary conditions are \emph{semi-admissible} if they satisfy
properties (\ref{difference}) and (\ref{bndrycon}). If $\mathbf{f}$
is in $L^2(\Omega)$ and (\ref{sys1}) is a symmetric
positive equation having semi-admissible boundary conditions, then (\ref{sys1})
possesses a weak solution in the ordinary sense: a vector
$\mathbf{u} \in L^2(\Omega)$ such that
\[
\int_\Omega\left(L^\ast\mathbf{w}\right)\cdot\mathbf{u}\,d\Omega =
\int_\Omega\mathbf{w}\cdot\mathbf{f}\,d\Omega
\]
for all vectors $\mathbf{w}$ having continuously differentiable
components and satisfying (\ref{adj}) (\cite{F}, Theorem 4.1).

Writing the higher-order terms of eqs.\ (\ref{sys2}), (\ref{sys3})
in the form
\begin{equation}\label{L}
L\textbf{u} =\left(%
\begin{array}{cc}
  K\left(x,y\right) & 0 \\
  0 & -1 \\
\end{array}%
\right)\left(%
\begin{array}{c}
  u_1 \\
  u_2 \\
\end{array}%
\right)_x + \left(%
\begin{array}{cc}
  0 & 1 \\
  1 & 0 \\
\end{array}%
\right)\left(%
\begin{array}{c}
  u_1 \\
  u_2 \\
\end{array}%
\right)_y,
\end{equation}
we will derive admissible boundary conditions for the system
(\ref{sys1})-(\ref{sys3}).

Slightly generalizing the type-change function of Sec.\ 2, we
choose $K\left(x,y\right)=x-\sigma(y),$ where $\sigma(y) \geq 0$
is a continuously differentiable function of its argument
satisfying (\emph{c.f.} \cite{O1})
\begin{equation}\label{sig1}
    \sigma(0)=\sigma'(0)=0,
\end{equation}

\begin{equation}\label{sig2}
    \sigma'(y)\geq 0\,\forall y \geq 0,
\end{equation}
and
\begin{equation}\label{sig3}
    \sigma'(y)\leq 0\,\forall y \leq 0.
\end{equation}
Let the operator $L$ in (\ref{sys1}) be given by
\begin{equation}\label{sys21}
    \left( L\mathbf{u}\right) _{1}=\left[ x-\sigma(y) \right] u_{1x} +
u_{2y}+\kappa_1u_1+\kappa_2u_2,
\end{equation}
\begin{equation}\label{sys31}
    \left( L\mathbf{u}\right) _{2}=u_{1y}-u_{2x},
\end{equation}
where $\kappa_1$ and $\kappa_2$ are constants. We note that equations of this kind satisfy all the physical requirements that originally led to the selection of eq.\ (\ref{coldpl1}) in \cite{W1} as a model for cold plasma waves.

By the \emph{elliptic} portion $\partial\Omega^+$ of the boundary
we mean points $\left(\tilde x, \tilde y\right)$ of the domain
boundary on which the type-change function $K\left(\tilde x,
\tilde y\right)$ is positive and by the \emph{hyperbolic} portion
$\partial\Omega^-,$ boundary points for which the type-change
function is negative. The \emph{sonic} portion of the boundary
consists of boundary points on which the type-change function
vanishes.

In this section we prove a revision and extension of \cite{O2},
Theorem 9:

\bigskip

\begin{theorem}\label{Theorem3} Let $\Omega$ be a bounded, connected domain
of $\mathbb{R}^2$ having $C^2$ boundary $\partial\Omega.$ Let $\partial\Omega^+_1$
be a (possibly empty and not necessarily proper) subset of
$\partial\Omega^+.$ Let inequality (\ref{starlike1}) be satisfied on
$\partial\Omega^+\backslash\partial\Omega_1^+.$ On
$\partial\Omega_1^+$ let
\begin{equation}\label{starlike2}
    bn_1+cn_2 \leq 0
\end{equation}
and on $\partial\Omega\backslash\partial\Omega^+,$ let
\begin{equation}\label{starlike3}
    -bn_1+cn_2 \geq 0.
\end{equation}
Let $b(x,y)$ and $c(x,y)$ satisfy
\begin{equation}\label{Q0}
    b^2+c^2K\ne 0
\end{equation}
on $\Omega,$ with neither $b$ nor $c$ vanishing on $\Omega^+,$ and the inequalities:
\begin{equation}\label{Q1}
    2b\kappa_1-b_xK-b+c_yK-c\sigma'(y)>0 \mbox{ in } \Omega;
\end{equation}
\begin{eqnarray}\label{Q2}
\left(2b\kappa_1-b_xK-b+c_yK-c\sigma'(y)\right)\left(2c\kappa_2+b_x-c_y\right)
\nonumber\\
    -\left(b\kappa_2+c\kappa_1-c_xK-c-b_y\right)^2>0 \mbox{ in } \Omega;
\end{eqnarray}
\begin{equation}\label{bQ2}
    K\left(bn_1-cn_2\right)^2+\left(cKn_1+bn_2\right)^2\leq 0 \mbox{ on } \partial\Omega\backslash\partial
    \Omega^+.
\end{equation}
Let $L$ be given by (\ref{sys21}), (\ref{sys31}). Let the Dirichlet
condition
\begin{equation}\label{dirichlet}
    -u_1n_2+u_2n_1=0
\end{equation}
be satisfied on $\partial\Omega^+\backslash\partial\Omega^+_1$ and
let the Neumann condition
\begin{equation}\label{neumann}
    Ku_1n_1+u_2n_2=0
\end{equation}
be satisfied on $\partial\Omega^+_1.$ Then eqs.\ (\ref{sys1}),
(\ref{sys21}), (\ref{sys31}) possess a strong solution on $\Omega$
for every $\mathbf f\in L^2(\Omega).$
\end{theorem}

\emph{Proof}. Multiply both sides of eqs.\ (\ref{sys1}),
(\ref{sys21}), (\ref{sys31}) by the matrix
\begin{equation}\label{multiplier}
       E=\left(%
\begin{array}{cc}
  b & -cK \\
  c & b \\
\end{array}%
\right).
\end{equation}
Condition (\ref{Q0}) implies that $E$ is invertible on $\Omega,$
and conditions (\ref{Q1}), (\ref{Q2}) imply that the resulting
system is symmetric positive.

For all points $\left(\tilde x, \tilde y\right)\in\partial\Omega,$
decompose the matrix
\[
    \beta\left(\tilde x, \tilde y\right)=\left(%
\begin{array}{cc}
  K\left(bn_1-cn_2\right) & cKn_1+bn_2 \\
  cKn_1+bn_2 & -\left(bn_1-cn_2\right) \\
\end{array}%
\right)
\]
into a matrix sum having the form $\beta = \beta_++\beta_-.$

On $\partial\Omega^+\backslash\partial\Omega^+_1,$ decompose
$\beta$ into the submatrices
\[
\beta_+=\left(%
\begin{array}{cc}
  Kbn_1 & bn_2 \\
  Kcn_1 & cn_2 \\
\end{array}%
\right)
\]
and
\[
\beta_-=\left(%
\begin{array}{cc}
  -Kcn_2 & Kcn_1 \\
  bn_2 & -bn_1 \\
\end{array}%
\right).
\]
Then $\beta_-\mathbf{u}=0$ under boundary condition
(\ref{dirichlet}). We have
\[
\mu^\ast=\left(bn_1+cn_2\right)\left(%
\begin{array}{cc}
  K & 0 \\
  0 & 1 \\
\end{array}%
\right),
\]
so condition (\ref{starlike1}) implies that the Dirichlet
condition (\ref{dirichlet}) is semi-admissible on
$\partial\Omega\backslash\partial\Omega^+_1.$

On $\partial\Omega_1^+,$ choose
\[
\beta_+=\left(%
\begin{array}{cc}
  -Kcn_2 & Kcn_1 \\
  bn_2 & -bn_1 \\
\end{array}%
\right)
\]
and
\[
\beta_-=\left(%
\begin{array}{cc}
  Kbn_1 & bn_2 \\
  Kcn_1 & cn_2 \\
\end{array}%
\right).
\]
Then $\beta_-\mathbf{u}=0$ under the Neumann boundary condition
(\ref{neumann}), and
\[
\mu^\ast=-\left(bn_1+cn_2\right)\left(%
\begin{array}{cc}
  K & 0 \\
  0 & 1 \\
\end{array}%
\right)
\]
is positive semi-definite under condition (\ref{starlike2}).

On $\partial\Omega\backslash\partial\Omega^+,$ choose
$\beta_+=\beta$ and take $\beta_-$ to be the zero matrix. Then
$\mu=\mu^\ast=\beta$ and
\[
\mu_{11}=K\left(bn_1-cn_2\right).
\]
Because $\mu_{11}$ is non-negative by (\ref{starlike3}),
$\mu^\ast$ is positive semi-definite by inequality (\ref{bQ2}),
and no conditions need be imposed outside the elliptic portion of
the boundary.

This yields semi-admissibility. We now prove admissibility.

On $\partial\Omega^+\backslash\partial\Omega^+_1$ the null space
of $\beta_-$ is composed of vectors satisfying the Dirichlet
condition (\ref{dirichlet}), which is imposed on that boundary
arc. The null space of $\beta_+$ is composed of vectors satisfying
the adjoint condition (\ref{neumann}). On $\partial\Omega_1^+,$
this relation is reversed. In order to show that the direct sum of
these null spaces spans the two-dimensional space
$\mathcal{V}_{|\partial\Omega^+},$ it is sufficient to show that
the set
\[
\left\lbrace\left(%
\begin{array}{c}
  1 \\
  n_2/n_1\\
\end{array}%
\right),\left(%
\begin{array}{c}
  1 \\
  -Kn_1/n_2  \\
\end{array}%
\right)\right\rbrace
\]
is linearly independent there. Setting
\[
c_1\left(%
\begin{array}{c}
  1 \\
  n_2/n_1\\
\end{array}%
\right)+c_2\left(%
\begin{array}{c}
  1 \\
  -Kn_1/n_2  \\
\end{array}%
\right)=\left(%
\begin{array}{c}
  0 \\
  0  \\
\end{array}%
\right),
\]
we find that $c_1=-c_2$ and
\begin{equation}\label{LI}
    -c_2\left(\frac{n_2^2+Kn_1^2}{n_1n_2}\right)=0.
\end{equation}
Equation (\ref{LI}) can only be satisfied on the elliptic boundary
if $c_2=0,$ implying that $c_1=0.$ Thus the direct sum of the null
spaces of $\beta_{\pm}$ on $\partial\Omega^+$ is linearly
independent and must span $\mathcal{V}$ over that portion of the
boundary.

On $\partial\Omega\backslash\partial\Omega^+,$ the null space of
$\beta_-$ contains every 2-vector and the null space of $\beta_+$
contains only the zero vector; so on that boundary arc, their
direct sum spans $\mathcal{V}.$

On $\partial\Omega^+\backslash\partial\Omega_1^+,$ the range
$\mathfrak{R}_+$ of $\beta_+$ is the subset of the range
$\mathfrak{R}$ of $\beta$ for which
\begin{equation}\label{R1}
    v_2n_1-v_1n_2=0
\end{equation}
for $\left(v_1,v_2\right)\in \mathcal{V};$ the range
$\mathfrak{R}_-$ of $\beta_-$ is the subset of $\mathfrak R$ for
which
\begin{equation}\label{R2}
    Kv_1n_1+v_2n_2=0
\end{equation}
for $\left(v_1,v_2\right)\in \mathcal{V}.$ Analogous assertions
hold on $\partial\Omega_1^+,$ in which the ranges of
$\mathfrak{R}_+$ and $\mathfrak{R}_-$ are interchanged. Because if
$n_1$ and $n_2$ are not simultaneously zero the system (\ref{R1}),
(\ref{R2}) has only the trivial solution $v_2=v_1=0$ on
$\partial\Omega^+,$ we conclude that
$\mathfrak{R}_+\cap\mathfrak{R}_-=\lbrace 0 \rbrace$ on
$\partial\Omega^+.$

On $\partial\Omega\backslash\partial\Omega^+,$
$\mathfrak{R}_-=\lbrace 0\rbrace,$ so
$\mathfrak{R}_+\cap\mathfrak{R}_-=\lbrace 0 \rbrace$ trivially.

The invertibility of $E$ under condition (\ref{Q0}) completes the
proof of Theorem \ref{Theorem3}.

\bigskip

\textbf{Remarks}. \emph{i}) By taking $\partial\Omega_1^+$ to be
either the empty set or all of $\partial\Omega^+,$ Theorem 3.1 implies
the existence of strong solutions for either the open Dirichlet
problem or the open Neumann problem. Because only the open Dirichlet
problem was considered in Theorem 9 of \cite{O2}, Theorem
\ref{Theorem3} of this report extends that result to the open cases
of the Neumann and mixed Dirichlet-Neumann problems.

\emph{ii}) A misprint in eq.\ (45) of \cite{O2} has been corrected
in eq.\ (\ref{sys21}). In Theorem \ref{Theorem3}, condition
(\ref{Q0}) has been added to the list of hypotheses in Theorem 9 of
\cite{O2}, the redundant condition (57) removed, and an error in
eq.\ (59) corrected by eq.\ (\ref{bQ2}) of this report. The
assumption that the boundary is piecewise smooth, which was default
hypothesis in \cite{O2}, seems to be too weak in general for strong
solutions; see, however, \cite{LaP}, \cite{Ln}, and
\cite{S1}-\cite{So2}.

\emph{iii}) Only conditions (\ref{Q1}) and (\ref{Q2}) have anything
to do with the cold plasma model. Otherwise, Theorem \ref{Theorem3}
is about interpreting Friedrichs' theory in the context of a
collection of boundary arcs which are starlike with respect to a
corresponding collection of vector fields. For example, the argument
leading to eq.\ (\ref{LI}) suggests that the Tricomi problem is
strongly ill-posed under the hypotheses of the theorem, whatever the
type-change function $K.$ This is because in the Tricomi problem,
data are given on both the elliptic boundary and a characteristic
curve; but on characteristic curves, $K$ satisfies
\begin{equation}\label{char}
    K=-\frac{n_2^2}{n_1^2}.
\end{equation}
Substituting this equation into eq.\ (\ref{LI}), we find that the
equation is satisfied on characteristic curves without requiring
the constants $c_1$ and $c_2$ to be zero.

However, the theorem is less restrictive if the operator in
(\ref{sys1}) is given by
\begin{eqnarray}
    \left( L\mathbf{u}\right) _{1}=\left[ x-\sigma(y) \right] u_{1x} -
u_{2y}+\kappa_1u_1+\kappa_2u_2, \nonumber\\
\left( L\mathbf{u}\right) _{2}=-u_{1y}+u_{2x}, \label{PFeq}
\end{eqnarray}
where, again, $\kappa_1$ and $\kappa_2$ are constants. This
variant also arises in the cold plasma model (see \cite{MSW} and
\cite{PF}) and is analogous to the variant of the Tricomi
equation,
\[
yu_{xx}-u_{yy}=0,
\]
studied in various contexts by Friedrichs \cite{F}, Katsanis
\cite{K}, Sorokina \cite{So1}, \cite{So2}, and Didenko \cite{D}.
In that case, choose
\[
    E=\left(%
\begin{array}{cc}
  b & cK \\
  c & b \\
\end{array}%
\right).
\]
Obvious modifications of conditions (\ref{Q1}) and (\ref{Q2})
guarantee that the equation
\[
EL\mathbf{u}=E\mathbf{f}
\]
will be symmetric positive. Condition (\ref{Q0}) must be replaced
by the invertibility condition
\[
    b^2-c^2K\ne 0,
\]
which is restrictive on the subdomain $\Omega^+$ rather than on
$\Omega^-$ as in (\ref{Q0}). Most importantly, the discussion
leading to Table 1 of \cite{K} now applies, with only minor
changes, and one can obtain a long list of possible starlike
boundaries which result in strong solutions to suitably formulated
problems of Dirichlet or Neumann type. In particular, one can
formulate a Tricomi problem which is strongly well-posed.

\emph{iv}) The hypotheses of Theorem \ref{Theorem3} have a rather
formal appearance. We expect them to be harsh, as the known
singularity at the origin should dramatically restrict the kinds of
smoothness results that we can prove. But many of the conditions
have natural interpretations. For example, inequalities
(\ref{starlike1}), (\ref{starlike2}), and (\ref{starlike3}) are
satisfied whenever boundary arcs are starlike with respect to an
appropriate vector field, and (\ref{bQ2}) is always satisfied on the
characteristic boundary:

\begin{proposition}\label{Proposition4} Let $\Gamma$ be a characteristic curve
for eq.\ (\ref{sys1}), with the higher-order terms of the operator
$L$ satisfying (\ref{L}). Then the left-hand side of inequality
(\ref{bQ2}) is identically zero on $\Gamma.$
\end{proposition}

\emph{Proof}. We have, using eq.\ (\ref{char}),
\[
\left(cKn_1+bn_2\right)^2=c^2K^2n_1^2+2Kcbn_1n_2+b^2n_2^2
\]

\[
=-c^2K^2\frac{n_2^2}{K}+2Kcbn_1n_2-b^2Kn_1^2=-K\left(c^2n_2^2-2cbn_1n_2+b^2n_1^2\right)
\]

\[
=-K\left(cn_2-bn_1\right)^2.
\]
Substituting the extreme right-hand side of this equation into the
second term of (\ref{bQ2}) completes the proof.

\subsection{An explicit example}

A simple example which illustrates the hypotheses of Theorem
\ref{Theorem3} can be constructed for the special case
$\sigma(y)\equiv 0.$ In that case the system (\ref{sys1}),
(\ref{sys21}), (\ref{sys31}) can be reduced, by taking $u_1=u_x,$
$u_2=u_y,$ and $\mathbf{f}=0,$ to the \emph{Cinquini-Cibrario
equation} \cite{C1}
\begin{equation}\label{cc}
    xu_{xx}+u_{yy} + \mbox{ lower-order terms }=0.
\end{equation}
In the context of the cold plasma model, this case corresponds to
a resonance curve which is collinear with a flux line. In such
situations eq.\ (\ref{coldpl1}) can be replaced by an ordinary
differential equation, as was discussed at the end of Sec.\ 2, so
this choice is of little direct interest for the cold plasma
model. But the Cinquini-Cibrario equation is interesting in its
own right, in connection with normal forms for second-order linear
elliptic-hyperbolic equations \cite{Bi}, \cite{CC1}. Moreover, polar forms of
eq.\ (\ref{cc}) arise in models of atmospheric and space plasmas
$-$ compare eq. (9) of \cite{J}, eq.\ (16) of \cite{T}, and eq.\
(19a) of \cite{TS}, with Sec.\ 3 of \cite{C2}. (Such plasmas are
cold, but in a relative rather than absolute sense.)

In addition to taking $\sigma(y)\equiv 0,$ choose $\kappa_1=1;$
$\kappa_2=0;$ $b=x+M,$ where $M$ is a positive constant which is
assumed to be large in comparison with all other parameters of the
problem $-$ in particular, $b>0\,\forall x\in\overline\Omega$;
$c=\epsilon y,$ where $\epsilon$ is a small positive constant;
$\left(n_1,n_2\right)=\left(dy/ds,-dx/ds\right),$ where $s$ is arc
length on the boundary. Inequalities (\ref{Q0})-(\ref{Q2}) are
satisfied for $M$ sufficiently large.

Let the hyperbolic region $\Omega^-$ be bounded by intersecting
characteristic curves originating on the sonic line. Condition
(\ref{bQ2}) is satisfied on $\partial\Omega^-$ by Proposition
\ref{Proposition4}. Condition (\ref{starlike3}) is satisfied for $M$
sufficiently large. As a concrete example, let $\partial\Omega^- =
\Gamma^-\cup\Gamma^+,$ where
\[
\Gamma^\pm = \left\{\left(x,y\right)\in\Omega^-|y= \pm
2\left(\sqrt{-x}-2\right)\right\}.
\]
These curves intersect at the point $\left(-4,0\right).$ Their
intersection is not $C^2,$ but it can be easily ``smoothed out"
(by the addition of a small noncharacteristic curve connecting the
points $\left(-4+\delta_0,\pm\delta_1\right)$ for
$0<\delta_0,\delta_1<<1$) without violating either of the
governing inequalities. Let the elliptic boundary
$\partial\Omega^+$ be a smooth convex curve, symmetric about the
$x$-axis, with endpoints at $\left(0,\pm 4\right)$ on the sonic
line. Let the disconnected subset $\partial\Omega_1^+$ of
$\partial\Omega^+$ take the form of two small ``smoothing curves,"
on which the slope of the tangent line to $\partial\Omega^+$
changes sign in order to prevent a cusp at the two endpoints.
Inequality (\ref{starlike1}) is satisfied on
$\partial\Omega^+\backslash\Omega^+_1$ and, again assuming that
$M$ is sufficiently large, inequality (\ref{starlike2}) is
satisfied on the two smoothing curves comprising
$\partial\Omega_1^+.$

The domain $\Omega$ of this construction is identical to the
domain illustrated in Fig.\ 2 on p.\ 277 of \cite{C2}, except that
the cusps at points $R,$ $M,$ and $N$ of that figure are smoothed
out in $\Omega$ near the points $\left(-4,0\right)$ and
$\left(0,\pm 4\right).$

Theorem \ref{Theorem3} implies that strong solutions to a
homogeneous mixed Dirichlet-Neumann problem for the first-order
inhomogeneous form of the Cinquini-Cibrario equation exist on this
natural class of domains for any square-integrable forcing function.

\subsection{Symmetric positive operators on domains having multiply starlike boundaries}

On the basis of the observations in Remarks \emph{iii)} and \emph{iv)}, we
reformulate Theorem \ref{Theorem3} for an arbitrary type-change
function $-$ that is, for a smooth function $K\left(x,y\right)$
which is positive on all points in a subset $\Omega^+$ of $\Omega,$
negative on all points of a subset $\Omega^-$ of $\Omega,$ and zero
on a smooth curve $\mathfrak{K}\in\Omega$ separating $\Omega^+$ and
$\Omega^-,$ where
\[
\Omega = \Omega^+\cup\Omega^-\cup\mathfrak{K}.
\]
The proof of Theorem \ref{Theorem3} will also prove:

\begin{corollary}\label{Corollary5} Let $\Omega$ be a bounded, connected
domain of $\mathbb{R}^2$ having $C^2$ boundary $\partial\Omega,$
oriented in a counterclockwise direction. Let $\partial\Omega^+_1$
be a (possibly empty and not necessarily proper) subset of
$\partial\Omega^+.$ Suppose that $EL$ is a symmetric positive
operator, where $L$ satisfies (\ref{L}) (with the possible addition
of lower-order terms) and $E$ satisfies (\ref{multiplier}) with
condition (\ref{Q0}). Let
$\partial\Omega^+\backslash\partial\Omega_1^+$ be starlike with
respect to the vector field $V^+ = -\left(b(x,y),c(x,y)\right);$ let
$\partial\Omega_1^+$ be starlike with respect to the vector field
$V^+_1=\left(b(x,y),c(x,y)\right);$ let
$\partial\Omega\backslash\partial\Omega^+$ be starlike with respect
to the vector field $V^-= \left(b(x,y),-c(x,y)\right).$ Let the
union $\partial\Omega\backslash\partial\Omega^+$ of the parabolic
and hyperbolic boundaries be subcharacteristic in the sense of
(\ref{bQ2}). Then the mixed boundary value problem given by eqs.\
(\ref{sys1})-(\ref{sys3}), with condition (\ref{dirichlet})
satisfied on $\partial\Omega^+\backslash\partial\Omega^+_1$ and
condition (\ref{neumann}) satisfied on $\partial\Omega^+_1,$
possesses a strong solution for every $\mathbf f \in L^2(\Omega).$
\end{corollary}

\bigskip

\textbf{Acknowledgments}

\bigskip

I am grateful to Daniela Lupo and Yuxi Zheng for discussion of condition (\ref{limit}).

\end{document}